\documentclass[manuscript,nonacm,format=acmtog]{acmart}
\AtBeginDocument{%
  }

\setcopyright{none}

\usepackage{xcolor}
\usepackage{algorithm}
\usepackage{algpseudocode}
\usepackage{tcolorbox}
\usepackage{amsmath}
\usepackage{url}
\usepackage[T1]{fontenc}
\usepackage{subcaption}
\usepackage{setspace}
\usepackage[bottom]{footmisc}
\usepackage[outline]{contour}
\graphicspath{ {./figures/} }
\DeclareGraphicsExtensions{.pdf,.jpeg,.png}
\authorsaddresses{}
\begin{document}

\newcommand{\mitransient}{\texttt{mitransient}}
\title{\mitransient: Transient light transport in Mitsuba 3}

\author{Diego Royo}
\email{droyo@unizar.es}
\orcid{0000-0001-6880-322X}
\author{Jorge Garcia-Pueyo}
\email{jorge.garciap@unizar.es}
\affiliation{%
  \institution{Universidad de Zaragoza - I3A}
  \city{Zaragoza}
  \state{Aragon}
  \country{Spain}
}

\author{Miguel Crespo}
\affiliation{%
  \institution{École Polytechnique Fédérale de Lausanne (EPFL)}
  \city{Lausanne}
  \country{Switzerland}}
\email{miguel.crespo@epfl.ch}

\author{Guillermo Enguita}
\email{genguita@unizar.es}
\author{Oscar Pueyo-Ciutad}
\email{o.pueyo@unizar.es}
\author{Diego Bielsa}
\affiliation{%
  \institution{Universidad de Zaragoza - I3A}
  \city{Zaragoza}
  \state{Aragon}
  \country{Spain}
}

\renewcommand{\shortauthors}{Diego Royo, Jorge García-Pueyo, Miguel Crespo, Óscar Pueyo-Ciutad, Guillermo Enguita and Diego Bielsa}
% \renewcommand{\shortauthors}{Royo et al.}

\begin{abstract}
  \mitransient\ is a light transport simulation tool that extends Mitsuba 3 with support for time-resolved simulations. In essence, \mitransient\ extends conventional rendering by adding a temporal dimension which accounts for the time of flight of light. This allows rapid prototyping of novel transient imaging systems without the need of costly or difficult-to-operate hardware. Our code is trivially easy to install through \texttt{pip}, and consists of Python modules that can run both in CPU and GPU by leveraging the JIT capabilities of Mitsuba 3. It provides physically-based simulations of complex phenomena, including a wide variety of realistic materials and participating media such as fog or smoke. In addition, we extend Mitsuba 3's functionality to support time-resolved polarization tracking of light and transient differentiable rendering. Finally, we also include tools that simplify the use of our simulations for non-line-of-sight imaging, enabling realistic scene setups with capture noise to be simulated in just seconds of minutes. Altogether, we hope that \mitransient\ will support the research community in developing novel algorithms for transient imaging.
\end{abstract}

%!TEX root = main.tex
\newcommand{\fref}[1]{Fig.~\ref{#1}}
\newcommand{\ffref}[2]{Figs.~\ref{#1}~and~\ref{#2}}
\newcommand{\tref}[1]{Tab.~\ref{#1}}
\newcommand{\eref}[1]{Eq.~\ref{#1}}
\newcommand{\eeref}[2]{Eqs.~\ref{#1} and \ref{#2}}
\newcommand{\erref}[2]{Eqs.~\ref{#1} to \ref{#2}}
\newcommand{\eeeref}[3]{Eqs.~\ref{#1}, \ref{#2}, and \ref{#3}}
\newcommand{\cref}[1]{Chap.~\ref{#1}}
\newcommand{\ssref}[2]{Secs.~\ref{#1}~and~\ref{#2}}
\newcommand{\sref}[1]{Sec.~\ref{#1}}
\newcommand{\aref}[1]{Appx.~\ref{#1}}

\graphicspath{{figures}{figures-other}}

% Commenting tools
\definecolor{red}{rgb}{0.8,0,0}
\definecolor{purered}{rgb}{1,0,0}
\definecolor{darkred}{rgb}{0.6,0,0}
\definecolor{green}{rgb}{0.0,0.5,0}
\definecolor{blue}{rgb}{0,0,0.75}
\definecolor{lightblue}{rgb}{0.3,0.3,0.75}
\definecolor{darkblue}{rgb}{0,0,0.55}
\definecolor{orange}{rgb}{0.9,0.3,0.1}
\definecolor{purple}{rgb}{0.6,0.0,0.6}
\definecolor{cyan}{rgb}{0.0,0.7,0.7}
\definecolor{darkgray}{rgb}{0.4,0.4,0.4}
\definecolor{bronze}{rgb}{0.8, 0.5, 0.2}
\definecolor{dorange}{rgb}{0.75, 0.4, 0.0}
\definecolor{grenate}{rgb}{0.76, 0.12, 0.28}

\renewcommand{\comment}[3]{\leavevmode\textcolor{#1}{\emph{(\textbf{#2:} #3)}}}
\newcommand{\colortxt}[2]{\leavevmode\textcolor{#1}{#2}}
\newcommand{\diegor}[1]{\comment{blue}{DiegoR}{#1}}
\newcommand{\jorge}[1]{\comment{orange}{Jorge}{#1}}
\newcommand{\oscar}[1]{\comment{green}{Oscar}{#1}}
\newcommand{\guillermo}[1]{\comment{red}{Guillermo}{#1}}
\newcommand{\diegob}[1]{\comment{purple}{DiegoB}{#1}}

\newcommand{\old}[1]{{\leavevmode\color{darkred}{#1}}}
\newcommand{\iccvfinal}[1]{#1}
\newcommand{\iccv}[1]{#1}
\newcommand{\new}[1]{{{#1}}}
\newcommand{\txtdone}[1]{{\leavevmode\color{green}{#1}}}

\newcommand{\todo}{$\square$\xspace}%
\newcommand{\cmark}{\ding{51}}%
\newcommand{\xmark}{\ding{55}}%
\newcommand{\tododone}{\cmark\xspace}%
\newcommand{\todofail}{\xmark\xspace}%
\newcommand{\enumstrength}{S.\arabic*}%
\newcommand{\enumproblem}{P.\arabic*}%

\renewcommand{\shortcite}[1]{\cite{#1}}

%\newlist{todolist}{enumerate}{10}
%\setlist[todolist]{label=T.\arabic*}

% Usual terms
\newcommand*{\pmapping}[0]{\emph{photon mapping}}
\newcommand*{\Pmapping}[0]{\emph{Photon mapping}}
\newcommand*{\PMapping}[0]{\emph{Photon Mapping}}
\newcommand*{\pbeams}[0]{\emph{photon beams}}
\newcommand*{\Pbeams}[0]{\emph{Photon beams}}
\newcommand*{\PBeams}[0]{\emph{Photon Beams}}
\newcommand*{\Ptracing}[0]{\emph{Path tracing}}
\newcommand*{\VPTracing}[0]{\emph{Volumetric Path Tracing}}
\newcommand*{\PTracing}[0]{\emph{Path Tracing}}
\newcommand*{\ptracing}[0]{\emph{path tracing}}
\newcommand*{\Bptracing}[0]{\emph{Bidirectional path tracing}}
\newcommand*{\bptracing}[0]{\emph{bidirectional path tracing}}
\newcommand*{\Metropolis}[0]{\emph{Metropolis light transport}}
\newcommand*{\brdf}[0]{\emph{BRDF}}
\newcommand*{\ns}[0]{\mathrm{ns}}
% Usual math and physics terms

\newcommand{\Tr}{{T_r}}
\newcommand{\abs}{\mu_a} %absorption coefficient
\newcommand{\sca}{\mu_s} %scattering coefficient
\newcommand{\ext}{\mu_t} %extinction coefficient
\newcommand{\alb}{{_\Lambda}}
\newcommand{\scalb}{\alpha}
\newcommand{\crossscat}{\kappa_s}
\newcommand{\crossabs}{\kappa_a}
\newcommand*{\pf}[0]{\rho}
\newcommand*{\coefext}[0]{\sigma_t}
\newcommand*{\coefextx}[0]{\ext(x)}
\newcommand*{\coefabs}[0]{\sigma_a}
\newcommand*{\coefabsx}[0]{\abs(x)}
\newcommand*{\coefscat}[0]{\sigma_s}
\newcommand*{\coefscatx}[0]{\sca(x)}
\newcommand*{\mcoefext}[0]{$\coefext$}
\newcommand*{\mcoefextx}[0]{$\coefextx$}
\newcommand*{\mcoefabs}[0]{$\coefabs$}
\newcommand*{\mcoefabsx}[0]{$\coefabsx$}
\newcommand*{\mcoefscat}[0]{$\coefscat$}
\newcommand*{\mcoefscatx}[0]{$\coefscatx$}
\newcommand*{\phasefunction}[0]{p(x, \vec{\omega}^\prime, \vec{\omega})}
\newcommand*{\ior}[0]{\eta}
%Usual equation expressions
\newcommand{\diff}{\mathrm{d}}

\newcommand{\omegav}{\vec{\omega}}
\newcommand{\omegaout}{\omegav_o}
\newcommand{\omegain}{\omegav_i}
\newcommand{\momegain}{$\omegain$}
\newcommand{\ppwr}{\Phi}
\newcommand{\eqbreak}{\nonumber \\}
\newcommand{\x}{\ensuremath{\mathbf{x}}\xspace}
\newcommand{\y}{\ensuremath{\mathbf{y}}\xspace}
\newcommand{\z}{\ensuremath{\mathbf{z}}\xspace}
\newcommand{\rvec}{\vec{\ensuremath{\mathbf{r}}}\xspace}
\newcommand{\rlen}{\ensuremath{\mathrm{r}}\xspace}
\newcommand{\xpr}{\ensuremath{\mathbf{x}^\prime}\xspace}
\newcommand{\ym}{{\y}}
\newcommand{\zm}{{\z}}
\newcommand{\VRegion}{\ensuremath{\aleph}\xspace}
\newcommand{\ys}{\ensuremath{\y_{\! s}}\xspace}
\newcommand{\xw}{\ensuremath{\hat{\x}}\xspace}
\newcommand{\yw}{\ensuremath{\hat{\y}}\xspace}
\newcommand{\mpar}{r}
\newcommand{\pkern}{\Omega}
\newcommand{\optpath}{\Pi}

% Steady-state transport matrix formulation
\newcommand{\stransportImage}{\mathbf{i}}
\newcommand{\stransportMatrix}{\mathbf{T}}
\newcommand{\stransportMatrixTransient}{\mathbf{H}}
\newcommand{\stransportSources}{\mathbf{p}}

% Virtual steady-state transport matrix formulation
\newcommand{\svtransportImage}{\mathbf{i_v}}
\newcommand{\svtransportMatrix}{\stransportMatrix}
\newcommand{\svtransportSources}{\mathbf{p_v}}
\newcommand{\sdefocusMatrix}{\mathbf{D}}

% Phasor fields terms
\newcommand{\norm}[1]{\left\lvert#1\right\rvert}
\newcommand{\Norm}[1]{\left\lVert#1\right\rVert}
\newcommand{\pftime}{t}
\newcommand{\planeC}{S}
\newcommand{\planeP}{L}
\newcommand{\xp}{\mathbf{x}_l}
\newcommand{\vvec}{\vec{\mathbf{v}}}
\newcommand{\dxp}{\diff \xp}
\newcommand{\xc}{\mathbf{x}_s}
\newcommand{\dxc}{\diff \xc}
\renewcommand{\x}{\mathbf{x}}
\newcommand{\xs}{\mathbf{x}_{s}}
\newcommand{\nonEmpty}{\Gamma}
\newcommand{\maskNonEmpty}{\textbf{m}_{\nonEmpty}}
\newcommand{\phasor}{\mathcal{P}}
\newcommand{\gating}{{G}}%{\mathbf{G}}
\newcommand{\dist}{{D}}%{\mathbf{D}}
\newcommand{\distf}[1]{\dist\left(#1\right)}
\newcommand{\gatingf}[1]{\gating\left(#1\right)}
\newcommand{\phasorf}[1]{\phasor\left(#1\right)}
\newcommand{\phasorwf}[1]{\phasor_{\pfFreq}\!\left(#1\right)}
\newcommand{\phasorxt}{\phasorf{\x, \pftime}}
\newcommand{\phasorwxt}{\phasorwf{\x, \pftime}}
\newcommand{\phasorxpt}{\phasorf{\xp, \pftime}}
\newcommand{\phasorwxpt}{\phasorwf{\xp, \pftime}}
\newcommand{\phasorxct}{\phasorf{\xc, \pftime}}
\newcommand{\phasorwxct}{\phasorwf{\xc, \pftime}}
\newcommand{\pfImagingModel}{\Phi}
\newcommand{\ROI}{V}
\newcommand{\pfImage}{I}
\newcommand{\pfImageof}[2]{\pfImage_{#1}\left(#2\right)}
\newcommand{\pfImpulse}{\stransportMatrixTransient}
\newcommand{\pfImpulseFun}{\pfImpulse\left(\xp,\xc,\pftime\right)}
\newcommand{\tof}{\text{\pftime}}
\newcommand{\dtof}[1]{\textbf{\pftime}_d\left(#1\right)}
\newcommand{\pfThinLens}{\mathcal{L}}
\newcommand{\pfThinLensFun}[2]{\pfThinLens_{#1}\!\left(#2\right)}
\newcommand{\pfFreq}{\omega}

% Polarization
\def\bbeta{\boldsymbol{\beta}}
\def\S{\boldsymbol{s}}
\def\M{\boldsymbol{M}}

% Other
\newcommand{\pathseq}[1]{\left\langle#1\right\rangle}

% For pseudocode
\tcbset{
  mylisting/.style={
    colback=gray!10,
    colframe=gray!50,
    boxrule=0.5pt,
    arc=3pt,
    outer arc=3pt,
    left=4pt,
    right=4pt,
    top=4pt,
    bottom=4pt,
  }
}

% Indents comments properly
\makeatletter
\newcommand{\CommentLine}[1]{%
  \Statex \hskip\ALG@thistlm {\color{darkgray}\# #1}%
}
\makeatother

% Remove last line of algorithms
\makeatletter
\newcommand\fs@nobottomruled{\def\@fs@cfont{\bfseries}\let\@fs@capt\floatc@ruled
  \def\@fs@pre{\hrule height.8pt depth0pt \kern2pt}%
  \def\@fs@post{}% Formerly \def\@fs@post{\kern2pt\hrule\relax}%
  \def\@fs@mid{\kern2pt\hrule\kern2pt}%
  \let\@fs@iftopcapt\iftrue}
\makeatother
\floatstyle{nobottomruled}
\restylefloat{algorithm}

\makeatletter
\algnewcommand{\linedots}{\Statex \hskip\ALG@thistlm $\ldots$}
\makeatother

% % ---- Force monospace font inside algorithms ----
% \makeatletter
% \renewcommand{\ALG@beginalgorithmic}{\ttfamily}
% \algrenewcommand\alglinenumber[1]{\texttt{\color{gray}#1}} % line numbers gray monospace
% \makeatother

% \newcommand{\CommentLine}[1]{\textcolor{darkgray}{\# #1}}
% Definir macro para 'def' en rojo y negrita
\algrenewcommand\algorithmicfunction{\textbf{def}} % <-- functions
\algrenewcommand\algorithmicprocedure{\textbf{def}} % (if you use Procedure instead of Function)

% Python-like control structures
\algrenewcommand\algorithmicwhile{\textbf{while}}
\algrenewcommand\algorithmicdo{:}
\algrenewcommand\algorithmicif{\textbf{if}}
\algrenewcommand\algorithmicthen{:}
\algrenewcommand\algorithmicelse{\textbf{else:}}
\algrenewcommand\algorithmicforall{\textbf{for}}
\algrenewcommand\algorithmicreturn{\textbf{return}}

% Remove the "end" statements
\algtext*{EndWhile}
\algtext*{EndIf}
\algtext*{EndFor}
\algtext*{EndFunction}
\algtext*{EndProcedure}

% Add ":" after function declaration
\algrenewtext{Function}[2]{\algorithmicfunction\ \textcolor{grenate}{\textbf{#1}}~(#2):}
\algrenewtext{Procedure}[2]{\algorithmicprocedure\ \textcolor{grenate}{\textbf{#1}}~(#2):}




\begin{teaserfigure}
    \centering
    % \captionsetup{skip=-6pt}
    \includegraphics[width=\textwidth]{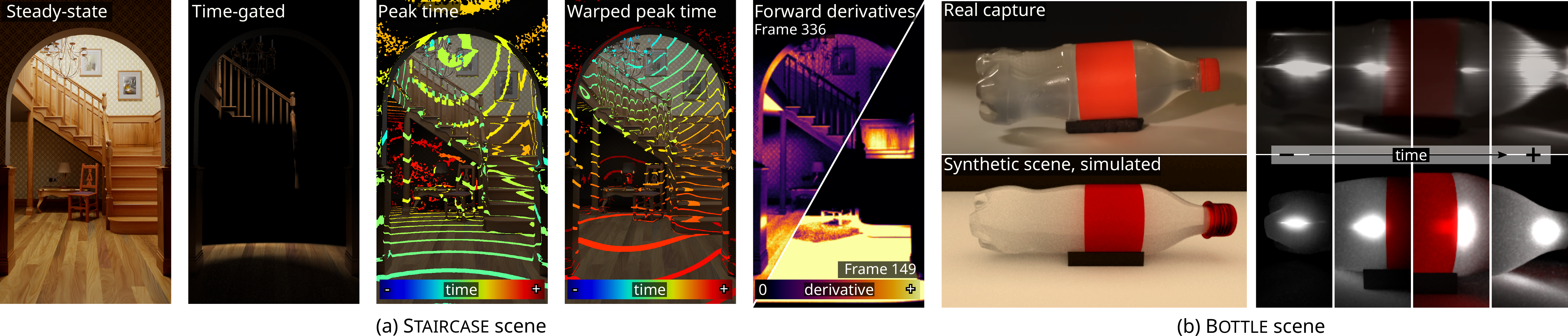}
    % \def\svgwidth{\textwidth}
    % \begin{small}
    %     \input{figures/fig1-v2-test.pdf_tex}
    % \end{small}
    \caption{\mitransient\ can simulate a wide variety of time-resolved and time-gated capture devices, with support for complex materials, participating media, polarization tracking, differentiable rendering and non-line-of-sight imaging. Some examples: \textbf{(a)} \textsc{Staircase} scene simulated using \texttt{\textbf{transient\_path}} integrator. From left to right: conventional camera, time-gated camera, peak time visualization, warped peak time visualization, and transient forward derivatives (with respect to the material of the floor). \textbf{(b)} \textsc{Bottle} scene filled with water and milk, being illuminated by a laser from its bottom. Up: real scene and frames of the propagation of light through the bottle captured with a streak camera. 
     Bottom: synthetic scene and frames computed by our \texttt{\textbf{transient\_volpath}}.}
    %\Description{figure description}
    \label{fig:cocacola}
\end{teaserfigure}

\maketitle

\section{Introduction}

Transient imaging aims to capture and analyze how light propagates and interacts through a scene with an ultra-high temporal resolution. 
This increase of information about the temporal domain provides several benefits that have led to numerous applications, such as visualization of light in motion~\cite{Jarabo2014}, vision through highly-scattering media~\cite{Luesia2022NLOSVision}, and inferring material properties from a distance.
One of the most prominent applications of transient imaging is the reconstruction of hidden objects from captured indirect illumination. This problem, known as non-line-of-sight~(NLOS) imaging, is especially relevant to our work.
Unfortunately, most methods that rely on transient imaging require expensive hardware, including ultra-fast cameras and pulsed illumination devices, which are difficult to calibrate and operate.

\emph{Transient light transport simulation} emerges as an alternative tool for developing and testing such systems, without hardware difficulties \cite{Jarabo2014}.
However, previous state-of-the-art systems for simulation are ad-hoc implementations for specific tasks (difficult to extend to other applications) or they lack novel optimized strategies (they are not as fast as they could be). Our rendering tool, \mitransient, solves these issues by being able to simulate different sorts of time-resolved sensing devices (including time-gated and transient cameras), with easy-to-extend modules written in Python that run on both the CPU and GPU by using vectorized JIT-compiled code from Mitsuba~3~\cite{Jakob2020DrJit}.

In \sref{sec:mitransient:software} we describe the software, documentation and usage of \mitransient. 
%Our software can be installed from PyPi\footnote{\url{https://pypi.org/project/mitransient/}}. Documentation is in Readthedocs\footnote{\url{https://mitransient.readthedocs.io}}, and issues/pull requests are on GitHub\footnote{\url{https://github.com/diegoroyo/mitransient/}}.
%
The rest of this article details the mathematical formulation behind transient rendering in \sref{sec:mitransient:transient_rendering}, where we also explain our visualization tools and output. And, in \sref{sec:mitransient:additional_features}, we showcase some of the capabilities of \mitransient:
\begin{itemize}
    % \item \textbf{Visualization} of transient renders (\sref{sec:mitransient:transient_rendering:visualization}).
    \item \textbf{Complex materials and volumetric scattering} (\sref{sec:mitransient:participating_media}).
    \item \textbf{Tracking and visualizing the polarization} of light in time (\sref{sec:mitransient:polarization}).
    \item \textbf{Rendering in frequency-space} for wave-based time-of-flight sensors (\sref{sec:mitransient:wave_rendering}).
    \item \textbf{Differentiable rendering} in the time domain. This allows you to compute gradients of transient renders with respect to scene parameters, enabling applications such as inverse rendering in the transient domain (\sref{sec:mitransient:differentiable}).
  \item \textbf{Simulation of non-line-of-sight (NLOS) setups} (\sref{sec:nlos}).
\end{itemize}

As this project is continuously evolving, we refer the reader to our website\footnote{\url{https://mitransient.readthedocs.io}} for up-to-date documentation, including custom installation instructions, tutorials, examples, and API specifications. Issues, pull requests and the full list of contributions is on GitHub\footnote{\url{https://github.com/diegoroyo/mitransient}}.

%%%%%%%%%%%%%%%%%%%%%%%%%%%%%%%%%%%%%%%%%%%%%%%%%%%%%%%%%%%%
%%%%%%%%%%%%%%%%%%%%%%%%%%%%%%%%%%%%%%%%%%%%%%%%%%%%%%%%%%%%
%%%%%%%%%%%%%%%%%%%%%%%%%%%%%%%%%%%%%%%%%%%%%%%%%%%%%%%%%%%%
\section{Software overview of \mitransient}
\label{sec:mitransient:software}

Our transient rendering tool, \mitransient, is an open-source Python package that extends the Mitsuba 3 physically-based renderer \cite{jakob2022mitsuba3}. We provide the package via PyPI\footnote{\url{https://pypi.org/project/mitransient/}}, and installation is as simple as \texttt{pip install mitransient}. As an example, the Python code below will compute a transient render of the \texttt{Cornell Box} scene and show a video of the result:

\begin{tcolorbox}[mylisting]
    \begin{algorithmic}[1]
        \State \textcolor{blue}{\textbf{import}} mitsuba \textcolor{blue}{\textbf{as}} mi
        \State mi.set\_variant(\textcolor{red}{\texttt{'llvm\_ad\_rgb'}})
        \State \textcolor{blue}{\textbf{import}} mitransient \textcolor{blue}{\textbf{as}} mitr
        \Statex
        \Comment{Or: load XML files with mi.load\_file()}
        \State scene \textcolor{blue}{\textbf{=}} mi.load\_dict(mitr.cornell\_box())
        \State img\_steady, img\_transient \textcolor{blue}{\textbf{=}} mi.render(scene, spp=\textcolor{magenta}{1024})
        \Statex
        \State img\_transient \textcolor{blue}{\textbf{=}} mitr.vis.tonemap\_transient(img\_transient)
        \Statex
        \Comment{Works best in IPython/Jupyter notebooks}
        \State mitr.vis.show\_video(img\_transient)
    \end{algorithmic}
\end{tcolorbox}

Thanks to Mitsuba's JIT-enabled variantes, the code can run on CPU (\texttt{llvm\_*} variants) or GPU (\texttt{cuda\_*} variants). Our codebase is modular and designed for ease of extension. We use also use Mitsuba's scene description XML files for convenient configuration.

%%%%%%%%%%%%%%%%%%%%%%%%%%%%%%%%%%%%%%%%%%%%%%%%%%%%%%%%%%%%%%%%%%%%%%%%%%%%%%%%%%%%%%%%%%%%%%%%%%%%%%%%
%%%%%%%%%%%%%%%%%%%%%%%%%%%%%%%%%%%%%%%%%%%%%%%%%%%%%%%%%%%%%%%%%%%%%%%%%%%%%%%%%%%%%%%%%%%%%%%%%%%%%%%%
%%%%%%%%%%%%%%%%%%%%%%%%%%%%%%%%%%%%%%%%%%%%%%%%%%%%%%%%%%%%%%%%%%%%%%%%%%%%%%%%%%%%%%%%%%%%%%%%%%%%%%%%
\section{Transient rendering}
\label{sec:mitransient:transient_rendering}

Light transport describes the process by which light is emitted from a source, interacts with the geometry and materials of the scene, and ultimately arrives at the camera sensor. This process is formally described through \emph{light paths} $\bar{\textbf{x}}$ defined as an ordered sequence of points:
\begin{equation}
    \bar{\textbf{x}} = \textbf{x}_0 \rightarrow \textbf{x}_1 \rightarrow \ldots \rightarrow \textbf{x}_k,
\end{equation}
where $\textbf{x}_0$ lies on an emitter, $\textbf{x}_k$ lies on the camera sensor, and $\textbf{x}_1 \ldots \textbf{x}_{k-1}$ are the intermediate surface or volume interaction points. The set of all possible light paths in a scene is called the \emph{path space} and is denoted with $\mathcal{X}$. For an image $I$, the value of a pixel $I(x, y, t)$ at coordinates $(x, y)$ and time $t$ is computed through the transient path integral formulation~\cite{veach1998robust, Jarabo2014}:
\begin{equation} \label{eq:mitransient:path-integration}
    I(x, y, t) = \int_{\mathcal{X}(x, y)} \mathcal{C}(\bar{\textbf{x}}, t) \, d\mu(\bar{\textbf{x}}),
\end{equation}
where $\mathcal{C}(\bar{\textbf{x}}, t)$ is the contribution of the path $\bar{\textbf{x}}$ at time $t$, and $\mathcal{X}(x, y) \subseteq \mathcal{X}$ is the subset of paths that reach that pixel. The differential measure $d\mu(\bar{\textbf{x}})$ denotes area integration for surface vertices and volume integration for vertices in participating media. The contribution of a path depends on its time of flight, which accounts for the time delays caused by light propagation from light to sensor. In free space, the time of flight $\text{tof}(\textbf{x}_i \rightarrow \textbf{x}_{i+1})$ required for light to travel from one point $\textbf{x}_i$ to another $\textbf{x}_{i+1}$ can be modeled simply as distance over the speed of light $c$, scaled by the index of refraction $\eta$ of the medium:
\begin{equation} \label{eq:mitransient:tof-one}
    \text{tof}(\textbf{x}_i \rightarrow \textbf{x}_{i+1}) = \Vert \textbf{x}_i - \textbf{x}_{i+1} \Vert \frac{\eta}{c}.
\end{equation}
The time of flight $\text{tof}(\bar{\textbf{x}})$ of a light path $\bar{\textbf{x}}$ accounts for all its segments:
\begin{equation}
    \text{tof}(\bar{\textbf{x}}) = \sum_{i=0}^{k-1} \text{tof}(\textbf{x}_i \rightarrow \textbf{x}_{i+1}).
\end{equation}
Next, we introduce two functions to model light emitted from the source and sensor sensitivity, which depend on time. First, the radiance $L_e(\mathbf{x}_0 \rightarrow \mathbf{x}_1, t)$ emitted from the source at $\mathbf{x}_0$ towards $\mathbf{x}_1$ at time $t$. By default, \mitransient\ models each source as a delta pulse, akin to an ultra-fast laser pulse:
\begin{equation}
    L_e(\mathbf{x}_0 \rightarrow \mathbf{x}_1, t) = \delta(t).
    \label{eq:mitransient:light-emission}
\end{equation}
Then, we define the sensitivity of the sensor $W_e(\textbf{x}_{k-1} \rightarrow \textbf{x}_k, t)$ located at $\textbf{x}_k$ for a ray of light coming from $\textbf{x}_{k-1}$. In the simplest case, $W_e$ is modeled as a rectangular function that is non-zero during the sensor exposure interval which spans between the times $b_\text{start}$ and $b_\text{end}$:
\begin{equation} \label{eq:mitransient:sensor-importance}
    W_e(\textbf{x}_{k-1} \rightarrow \textbf{x}_k, t) =
    \begin{cases}
        1, & b_\text{start} \leq t \leq b_\text{end}, \\
        0, & \text{otherwise},
    \end{cases}
\end{equation}
still, more complex functions can be modeled if necessary. Finally, we use the time of flight to model the contribution of a light path as the convolution between the emitted light $L_e$ and the scene impulse response $\mathfrak{T}$ over time, weighted by the sensitivity of the sensor $W_e$:
\begin{equation} \label{eq:mitransient:transient-path-integration}
    \begin{aligned}
        \mathcal{C}(\bar{\mathbf{x}}, t)
         & = \! \left[ \int_{-\infty}^{+\infty}\!\!
            L_e(\mathbf{x}_0 \!\!\rightarrow\! \mathbf{x}_1, \tau)\,
            \mathfrak{T}(\bar{\mathbf{x}}, t\!-\!\tau)\,
        \mathrm{d}\tau \right] W_e(\mathbf{x}_{k-1} \!\!\rightarrow\! \mathbf{x}_k, t)  \\
         & =  \left[ \left( L_e(\mathbf{x}_0 \!\!\rightarrow\! \mathbf{x}_1, \,\cdot\,)
            \ast_t \mathfrak{T}(\bar{\mathbf{x}}, \,\cdot\,)
            \right)(t) \right] W_e(\mathbf{x}_{k-1} \!\!\rightarrow\! \mathbf{x}_k, t).
    \end{aligned}
\end{equation}
Here $\mathfrak{T}(\bar{\mathbf{x}}, t)$ represents the temporal response of the path $\bar{\textbf{x}}$ to a pulse coming from $\textbf{x}_0$. At each interaction point $\textbf{x}_i$, $i \in \{1, ..., k-1\}$, light may experience a scattering delay due to absorption and re-emission (such as fluorescence or phosphorescence) or subsurface interactions. For all applications in this thesis, the scattering delays are negligible when compared to propagation times $\text{tof}(\textbf{x}_i \rightarrow \textbf{x}_{i+1})$. In this case, the temporal response of a pulse emitted at $t = 0$ will be another pulse that arrives at the sensor at $t = \text{tof}(\bar{\textbf{x}})$ as:
% only depends on the time of flight of the path $\bar{\textbf{x}}$ as $\mathfrak{T}(\bar{\mathbf{x}}, t) \propto \delta(t - \text{tof}(\bar{\mathbf{x}}))$. However one can also model materials with different scattering delays $\Delta t_i$ at each interaction point $\textbf{x}_i$, such as fluorescent materials. Thus, for the general case:
%
\begin{equation} \label{eq:mitransient:impulse-response}
    \mathfrak{T}(\bar{\mathbf{x}}, t) \!= \! \left[ \delta\left(t \!-\! \text{tof}(\bar{\mathbf{x}})\right) \!\prod_{i=1}^{k-1}\! \rho(\textbf{x}_i) \right]\! \left[ \prod_{i=0}^{k-1}\! G(\textbf{x}_i, \textbf{x}_{i+1}) V(\textbf{x}_i, \textbf{x}_{i+1}) \right]\!,
\end{equation}
where $\rho(\textbf{x}_i)$ is the scattering albedo at intersection point $\textbf{x}_i$, the geometry term $G(\textbf{x}_i, \textbf{x}_{i+1})$ accounts for foreshortening and distance falloff between points $\textbf{x}_i$ and $\textbf{x}_{i+1}$, and $V(\textbf{x}_i, \textbf{x}_{i+1})$ is the binary visibility term that is 1 if the two points are mutually visible, and 0 otherwise.
%This formulation is a bit different from the original proposed by Jarabo et al.~\cite{jarabo2014framework}, as we do not account for scattering delays at each interaction point $\textbf{x}_i$ yet in \mitransient. For all applications in this thesis, these delays are negligible when compared to propagation times. % Still, the full formulation is detailed later in \sref{sec:nlostr:transient_rendering}.

\paragraph{Monte Carlo estimation.} Given the large dimensionality of the path space $\mathcal{X}$, \emph{Monte Carlo} methods are typically used to numerically estimate the integral in \eref{eq:mitransient:path-integration} by randomly sampling $N$ light paths $\bar{\textbf{x}}_j$ from $\mathcal{X}$ according to a probability density function (PDF) $p(\bar{\textbf{x}}_j)$:
\begin{equation} \label{eq:mitransient:monte-carlo-estimator}
    I(x, y, t) \approx \frac{1}{N} \sum_{j=1}^N \frac{\mathcal{C}(\bar{\textbf{x}}_j, t)}{p(\bar{\textbf{x}}_j)}.
\end{equation}

\paragraph{Path reuse and histogramming.} Note that the contribution of a path $\mathcal{C}(\bar{\textbf{x}}, t)$ is non-zero only at $t = \text{tof}(\bar{\textbf{x}})$. Therefore, when estimating $I(x, y, t)$ for multiple time bins $t$, we can reuse the same $N$ sampled paths $\bar{\textbf{x}}$ for the same pixel $(x, y)$. This strategy, denoted as \emph{path reuse} \cite{Jarabo2014}, is implemented in \mitransient. The resulting paths $\bar{\textbf{x}}$ are histogrammed depending on their time of flight $\text{tof}(\bar{\textbf{x}})$. The specific bin depends on the start $b_\text{start}$ and end $b_\text{end}$ times of the whole histogram, along with the exposure time for each time bin $b_\text{width}$, with a resulting index of $\lfloor \left(\text{tof}(\bar{\textbf{x}}) - b_\text{start} \right) / b_\text{width} \rfloor$.

\paragraph{Camera warp and unwarping.} Before looking at any images, one needs to be aware of the effect of camera warp. At the time scales considered here, the time of flight of light from the scene to the camera is not negligible, meaning that the temporal ordering of events as they occur in the scene (\emph{world time}) differs from the ordering in which they are recorded by the sensor (\emph{camera time}). Concretely, this discrepancy comes from the time of flight $\text{tof}(\textbf{x}_{k-1} \rightarrow \textbf{x}_k)$ between the last interaction point at $\textbf{x}_{k-1}$ and the camera sensor at $\textbf{x}_k$. Following the work from \citet{Velten2013}, we may therefore apply \emph{camera unwarping} in our simulations by omitting this final time of flight, effectively allowing the camera to observe events in world time.% \droyo{Talk about how visualization is weird because you also need to account for the last bounce before the camera $\text{tof}(\textbf{x}_{k-1} \rightarrow \textbf{x}_k)$. How \emph{camera unwarping} is not accounting for that last bounce. Add a Figure of the staircase scene that shows these two.}

\fref{fig:cocacola}a showcases a transient render of the \texttt{Staircase} scene. All results have been computed using our \texttt{transient\_path} plugin from \mitransient, which implements a transient path tracing integrator. The \texttt{Staircase} scene is illuminated by a light on the ceiling. The light emits a pulse at $t=0$, and light travels from the top to the bottom of the image as time advances, bouncing on the geometry and eventually reaching the camera. The order of events slightly changes depending on whether we account for camera warp (denoted as \emph{camera time}) or not (denoted as \emph{world time}). We also include a \emph{peak time} visualization, inspired from work by \citet{Velten2013}, that shows the time of the maximum intensity for each pixel encoded by its color. Note how these peaks change from camera time to world time.

\section{Transient rendering extensions}
\label{sec:mitransient:additional_features}
This section highlights additional features of \mitransient\ that showcase the generality and extensibility of the framework.

\subsection{Complex materials and volumetric scattering}
\label{sec:mitransient:participating_media}

Notice that our system is capable of simulating conventional, time-gated, and transient cameras, including effects such as time unwarping~\cite{Velten2013}.
Furthermore, both RGB and spectral rendering are supported.
%
%\fref{fig:cocacola} shows multiple results from our transient algorithms in both transport regimes:
In \fref{fig:cocacola}a, we leverage the complex \textsc{Staircase} scene to demonstrate several computations that can be done with our system with light having multiple bounces inside of it.

\paragraph{Volumetric scattering.} Participating media such as fog, smoke or water influence light transport through absorption and scattering events that occur within the medium volume, and through a different index of refraction $\eta$ that affects light propagation speed (\eref{eq:mitransient:tof-one}). For an introduction to steady-state modeling of participating media, we refer the reader to the course by \citet{gutierrez2008scattering}. In \mitransient\ we extend our transient path tracing integrator to account for the time of flight of light inside participating media, denoted as the \texttt{transient\_prbvolpath} integrator.

In \fref{fig:cocacola}b we present a recreation of the iconic \textsc{Bottle} scene from Velten~et~al.~\cite{Velten2013}, including a bottle containing water with some drops of milk for added scattering, which we model using a rough plastic material containing a participating medium with coefficients similar to the physical ones.

\paragraph{Scattering delays.} Some materials, such as fluorescent or phosphorescent surfaces, do not scatter light instantaneously upon interaction. Instead, incoming photons may be absorbed and re-emitted after a non-negligible delay following a material-specific temporal distribution. Thus, the time of flight now needs to account for all the scattering delays along the path $\overline{\Delta t} = \{\Delta t_1, \Delta t_2, \ldots, \Delta t_{k-1}\}$, where $\Delta t_i$ is the scattering delay at interaction point $\textbf{x}_i$ as:
\begin{equation}
    \text{tof}(\bar{\textbf{x}}, \overline{\Delta t}) = \sum_{i=0}^{k-1} \text{tof}(\textbf{x}_i \rightarrow \textbf{x}_{i+1}) + \sum_{i=1}^{k-1} \Delta t_i.
\end{equation}
Finally, we extend \eref{eq:mitransient:impulse-response} so that $\mathfrak{T}(\textbf{x}, t)$ is not just another pulse, but a more complex signal that depends on the scattering delays along the path:
%
% \begin{equation}
%     \mathfrak{T}(\bar{\mathbf{x}}, t) = \left[ \int_\mathcal{T} \; \delta(t - \text{tof}(\bar{\textbf{x}}, \overline{\Delta t})) \; \prod_{i=1}^{k-1} \rho(\textbf{x}_i, \Delta t_i) \, \mathrm{d}\overline{\Delta t} \right] \left[ \prod_{i=0}^{k-1} G(\textbf{x}_i, \textbf{x}_{i+1}) V(\textbf{x}_i, \textbf{x}_{i+1}) \right],
% \end{equation}
\begin{equation}
\begin{aligned}
\mathfrak{T}(\bar{\mathbf{x}}, t)
&= \left[ \int_\mathcal{T} \delta\!\left(t - \text{tof}(\bar{\mathbf{x}}, \overline{\Delta t})\right)
   \prod_{i=1}^{k-1} \rho(\mathbf{x}_i, \Delta t_i)
   \, \mathrm{d}\overline{\Delta t} \right] \\
&\quad \left[
   \prod_{i=0}^{k-1} G(\mathbf{x}_i, \mathbf{x}_{i+1})
   V(\mathbf{x}_i, \mathbf{x}_{i+1})
\right].
\end{aligned}
\end{equation}
where $\rho(\textbf{x}_i, \Delta t_i)$ now also depends on the scattering delay $\Delta t_i$, and $\mathcal{T}$ is the set of all possible scattering delays along the path.

\paragraph{Spectral rendering.} Note that $I(x, y, t)$ can represent monochromatic intensity, a RGB triplet, or a full spectral distribution. For some contexts, rendering with a spectral distribution is necessary to capture phenomena such as material-specific reflectance and the sensor spectral response, which cannot be faithfully represented using RGB data. On the other hand, monochromatic rendering provides a computationally efficient alternative when only a single wavelength is required. In any case, \mitransient\ extends Mitsuba~3's spectral rendering capabilities to the transient domain, supporting both cases.
%%%%%%%%%%%%%%%%%%%%%%%%%%%%%%%%%%%%%%%%%%%%%%%%%%%%%%%%%%%%%%%%%%%%%%%%%%%%%%%%%%%%%%%%%%%%%%%%%%%%%%%%
%%%%%%%%%%%%%%%%%%%%%%%%%%%%%%%%%%%%%%%%%%%%%%%%%%%%%%%%%%%%%%%%%%%%%%%%%%%%%%%%%%%%%%%%%%%%%%%%%%%%%%%%
%%%%%%%%%%%%%%%%%%%%%%%%%%%%%%%%%%%%%%%%%%%%%%%%%%%%%%%%%%%%%%%%%%%%%%%%%%%%%%%%%%%%%%%%%%%%%%%%%%%%%%%%
\subsection{Transient polarization tracking}
\label{sec:mitransient:polarization}

\begin{figure}
    \centering
    \captionsetup{skip=-6pt}
    \def\svgwidth{\columnwidth}
    \begin{small}
        %% Creator: Inkscape 1.2.2 (b0a8486541, 2022-12-01), www.inkscape.org
%% Accompanies image file 'polarization.pdf' (pdf, eps, ps)
\begingroup%
  \makeatletter%
  \providecommand\color[2][]{%
    \errmessage{(Inkscape) Color is used for the text in Inkscape, but the package 'color.sty' is not loaded}%
    \renewcommand\color[2][]{}%
  }%
  \providecommand\transparent[1]{%
    \errmessage{(Inkscape) Transparency is used (non-zero) for the text in Inkscape, but the package 'transparent.sty' is not loaded}%
    \renewcommand\transparent[1]{}%
  }%
  \providecommand\rotatebox[2]{#2}%
  \newcommand*\fsize{\dimexpr\f@size pt\relax}%
  \newcommand*\lineheight[1]{\fontsize{\fsize}{#1\fsize}\selectfont}%
  \ifx\svgwidth\undefined%
    \setlength{\unitlength}{1089.38725509bp}%
    \ifx\svgscale\undefined%
      \relax%
    \else%
      \setlength{\unitlength}{\unitlength * \real{\svgscale}}%
    \fi%
  \else%
    \setlength{\unitlength}{\svgwidth}%
  \fi%
  \global\let\svgwidth\undefined%
  \global\let\svgscale\undefined%
  \makeatother%
  \begin{picture}(1,0.33043406)%
    \lineheight{1}%
    \setlength\tabcolsep{0pt}%
    \put(0,0){\includegraphics[width=\unitlength,page=1]{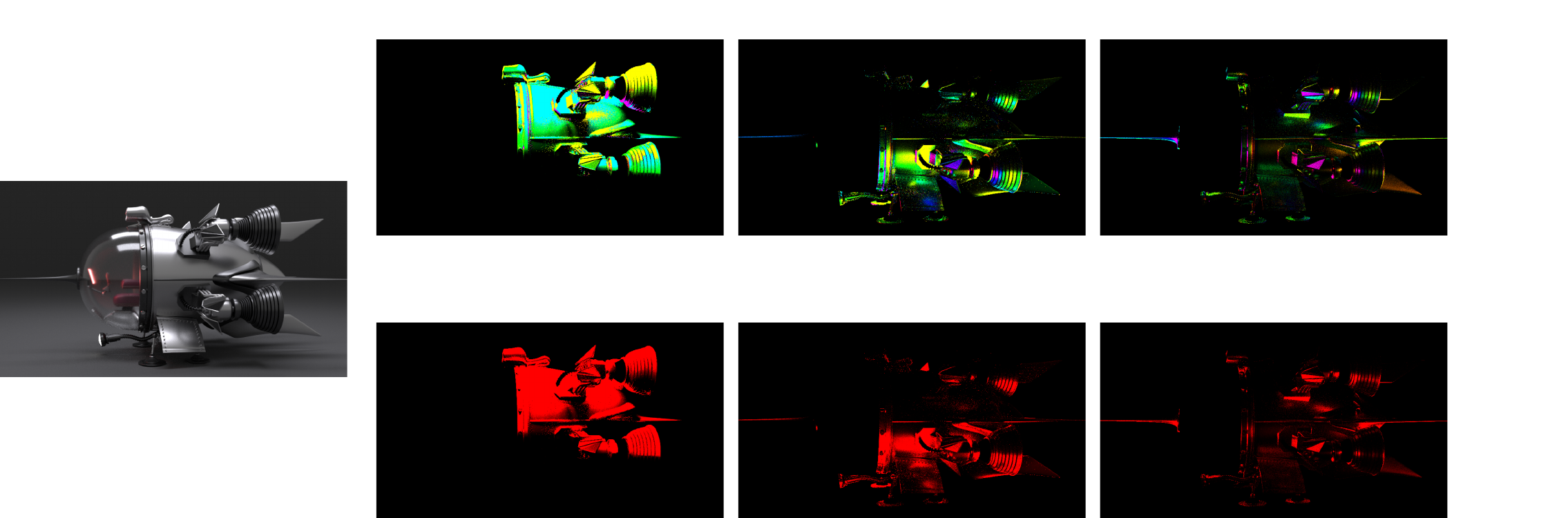}}%
    \put(0.11076248,0.22621417){\makebox(0,0)[t]{\lineheight{1.25}\smash{\begin{tabular}[t]{c}Steady state\end{tabular}}}}%
    \put(0.24012441,0.31619604){\makebox(0,0)[lt]{\lineheight{1.25}\smash{\begin{tabular}[t]{l}Angle of polarization $\xi$ over time\end{tabular}}}}%
    \put(0.24012441,0.13576777){\makebox(0,0)[lt]{\lineheight{1.25}\smash{\begin{tabular}[t]{l}Degree of polarization $\psi$ over time\end{tabular}}}}%
    \put(0,0){\includegraphics[width=\unitlength,page=2]{polarization.pdf}}%
    \put(0.92906752,0.23558958){\color[rgb]{1,1,1}\makebox(0,0)[lt]{\lineheight{1.25}\smash{\begin{tabular}[t]{l}$\xi$\end{tabular}}}}%
    \put(0.95084952,0.05264313){\makebox(0,0)[lt]{\lineheight{1.25}\smash{\begin{tabular}[t]{l}$\psi$\end{tabular}}}}%
    \put(0.95063798,0.00088969){\makebox(0,0)[lt]{\lineheight{1.25}\smash{\begin{tabular}[t]{l}0\end{tabular}}}}%
    \put(0.95061114,0.10617962){\makebox(0,0)[lt]{\lineheight{1.25}\smash{\begin{tabular}[t]{l}+\end{tabular}}}}%
  \end{picture}%
\endgroup%

    \end{small}
    \caption[Transient polarization in the \texttt{Spaceship} scene]{\textbf{Transient polarization in the \texttt{Spaceship} scene.} In \texttt{mitransient}, we support rendering the polarization state of transient light transport, represented by the Stokes vector of the radiance reaching the sensor. In the \texttt{Spaceship} scene, the polarization state at different frames varies due to the propagation delays between interactions of light in the spaceship. The angle of linear polarization uses a rainbow colormap and the degree of polarization a scale of reds, following the convention on polarized rendering \cite{WilkiePolvis10}.}
    \label{fig:polarized_render}
\end{figure}

The polarization of light refers to the orientation of the electric (and implicitly, the magnetic) field that composes the electromagnetic wave as it propagates through space. In unpolarized light, the electric field oscillates randomly in all directions perpendicular to the direction of propagation. In contrast, polarized light exhibits a preferred structure in these oscillations. The two fundamental forms of polarization are linear polarization (when the electric field oscillates in a fixed direction) and circular polarization (when the electric field rotates at a fixed rate). For each pixel $I(x, y, t)$ in an image, polarization is represented by a $4\times1$ Stokes vector per color channel. In a monochromatic context, this corresponds to:
\begin{equation}
    I(x, y, t) =
    \begin{bmatrix}
        S_0(x, y, t) \\[3pt]
        S_1(x, y, t) \\[3pt]
        S_2(x, y, t) \\[3pt]
        S_3(x, y, t)
    \end{bmatrix}.
    \!
    \begin{array}{@{}l@{}}
        =\text{total intensity of the light}              \\[3pt]
        =\text{linear polarization ($0^\circ/90^\circ$)} \\[3pt]
        =\text{linear polarization ($\pm 45^\circ$)}      \\[3pt]
        =\text{circular polarization.}
    \end{array}
\end{equation}
While conventional imaging systems measure only the total intensity reaching the sensor, specialized hardware can measure the incoming polarization state. This gives valuable information about the scene, as the polarization can heavily depend on the surface geometry and materials which light has interacted with along its path. For a comprehensive introduction to steady-state polarization tracking, we refer the reader to the work by \citet{WilkiePolarised2012}. Our integrators in \mitransient\ support tracking polarization through time: \fref{fig:polarized_render} shows an example in the \texttt{Spaceship} scene, showing the angle $\xi = \text{arctan}2(S_2, S_1)$ of linear polarization and degree of polarization $\psi = \sqrt{S_1^2 + S_2^2 + S_3^2}/S_0$ over time.

Being able to measure the polarization state of light transport at picosecond-scale resolution is a relatively recent development, and opens new avenues for computational imaging algorithms that leverage polarization information in the transient domain. For example, \citet{baek2022all} demonstrate that time-resolved polarimetric measurements can disentangle light path components such as direct reflections and subsurface scattering; \citet{scheuble2024polarization} use polarized LIDAR measurements to reconstruct large-scale scenes; and \citet{pueyo2024time}---whose work was developed using \mitransient---combine transient and polarization cues to improve non-line-of-sight imaging capabilities.
% The throughput value in polarized rendering is a Mueller matrix $\bbeta$ that tracks the polarization interaction each scattering event in each path vertex.
% Following \texttt{Mitsuba3} convention, each interaction is computed in local coordinates and transformed into its global coordinates through the \textbf{si.to\_world\_mueller()} function.
% Therefore, as we are tracking radiance in transient path tracing, the path's throughput is updated as $\bbeta = \bbeta \; \M_i$, where  $\M_i$ is the Mueller matrix of the scattering event at the $i$-th path vertex.

% NOTE(droyo): We can talk about this but I'd talk about the transient part, add t's to stuff.
% For Next Event Estimation (NEE), the captured polarization state by the sensor $\S_s$ is computed as
% $$
% \S_s = \bbeta \; \M_{\text{NEE}} \; \S_l,
% $$
% where $\M_{\text{NEE}}$ is the Mueller matrix of the scattering for NEE, and $\S_l$ is the Stokes vector representing the polarization state of light emitted by the light source.

%%%%%%%%%%%%%%%%%%%%%%%%%%%%%%%%%%%%%%%%%%%%%%%%%%%%%%%%%%%%%%%%%%%%%%%%%%%%%%%%%%%%%%%%%%%%%%%%%%%%%%%%
%%%%%%%%%%%%%%%%%%%%%%%%%%%%%%%%%%%%%%%%%%%%%%%%%%%%%%%%%%%%%%%%%%%%%%%%%%%%%%%%%%%%%%%%%%%%%%%%%%%%%%%%
%%%%%%%%%%%%%%%%%%%%%%%%%%%%%%%%%%%%%%%%%%%%%%%%%%%%%%%%%%%%%%%%%%%%%%%%%%%%%%%%%%%%%%%%%%%%%%%%%%%%%%%%
\subsection{Frequency-space rendering}
\label{sec:mitransient:wave_rendering}

Instead of directly measuring the arrival time for each individual photon, many practical imaging systems operate in the frequency domain. A notable example is Amplitude Modulated Continuous Wave (AMCW) LIDAR, commonly found in consumer time-of-flight cameras (e.g., Microsoft Kinect or autonomous vehicle sensors). These devices emit light modulated with a periodic waveform---typically a sine wave---and measure the phase shift
%and demodulation contrast
of the returning signal to estimate depth.
% and material properties
Similarly, fluorescence lifetime imaging microscopy analyzes the frequency response of samples to resolve lifetimes of fluorophores.

To simulate these systems, we compute the Fourier transform of the transient image $I(x, y, t)$, which yields the complex-valued image $\hat{I}(x, y, \Omega)$ at a specific angular frequency $\Omega$:
\begin{equation}
    \hat{I}(x, y, \Omega) = \mathcal{F} \left\{ I(x, y, t) \right\} = \int_{-\infty}^{+\infty} I(x, y, t^\prime) e^{-i \Omega t^\prime} \, \mathrm{d}t^\prime.
\end{equation}
where $\mathcal{F}$ represents a Fourier transform. A naive approach to computing this quantity is to render the full transient video $I(x, y, t)$ first, using the histogramming method described in \sref{sec:mitransient:transient_rendering}, and subsequently apply a discrete Fast Fourier Transform (FFT). This naive approach works well under certain conditions e.g., when one needs to compute the response for a very large number of frequencies $\Omega$. However, it requires rendering a sufficiently large temporal interval $[b_\text{start}, b_\text{end}]$ (\eref{eq:mitransient:sensor-importance}), and is prone to discretization errors as the histogram bin width $b_\text{width}$ needs to be sufficiently small to satisfy the Nyquist-Shannon sampling theorem.

%a sufficiently high temporal resolution in the histogram binning, or else it can lead to inaccurate results. This can be computationally expensive, especially when only a few frequency bins $\Omega$ are required. However, this method is computationally inefficient and prone to discretization errors. It requires rendering a high-resolution temporal video to satisfy the Nyquist-Shannon sampling theorem, demanding excessive memory storage and long exposure times $(t_\text{start}, t_\text{end})$ to ensure the full signal is captured.

\begin{figure}
    \centering
    \captionsetup{skip=-3pt}
    \def\svgwidth{\columnwidth}
    \begin{small}
        %% Creator: Inkscape 1.2.2 (b0a8486541, 2022-12-01), www.inkscape.org
%% Accompanies image file 'freqspace.pdf' (pdf, eps, ps)
\begingroup%
  \makeatletter%
  \providecommand\color[2][]{%
    \errmessage{(Inkscape) Color is used for the text in Inkscape, but the package 'color.sty' is not loaded}%
    \renewcommand\color[2][]{}%
  }%
  \providecommand\transparent[1]{%
    \errmessage{(Inkscape) Transparency is used (non-zero) for the text in Inkscape, but the package 'transparent.sty' is not loaded}%
    \renewcommand\transparent[1]{}%
  }%
  \providecommand\rotatebox[2]{#2}%
  \newcommand*\fsize{\dimexpr\f@size pt\relax}%
  \newcommand*\lineheight[1]{\fontsize{\fsize}{#1\fsize}\selectfont}%
  \ifx\svgwidth\undefined%
    \setlength{\unitlength}{1216.75556333bp}%
    \ifx\svgscale\undefined%
      \relax%
    \else%
      \setlength{\unitlength}{\unitlength * \real{\svgscale}}%
    \fi%
  \else%
    \setlength{\unitlength}{\svgwidth}%
  \fi%
  \global\let\svgwidth\undefined%
  \global\let\svgscale\undefined%
  \makeatother%
  \begin{picture}(1,0.25331452)%
    \lineheight{1}%
    \setlength\tabcolsep{0pt}%
    \put(0,0){\includegraphics[width=\unitlength,page=1]{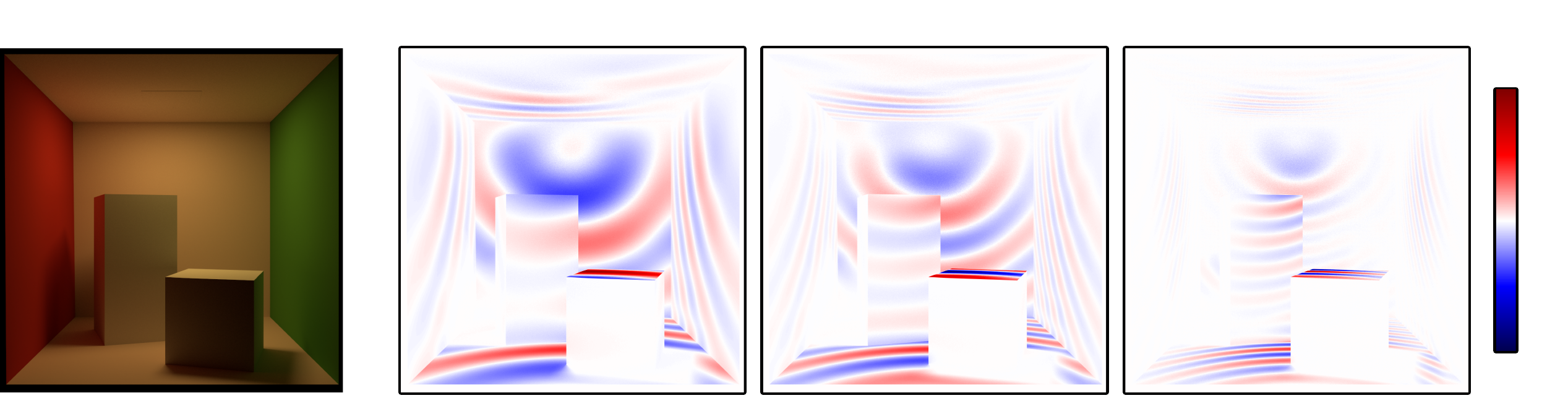}}%
    \put(0.10935311,0.234582){\makebox(0,0)[t]{\lineheight{1.25}\smash{\begin{tabular}[t]{c}Steady state\end{tabular}}}}%
    \put(0.36503756,0.23458193){\makebox(0,0)[t]{\lineheight{1.25}\smash{\begin{tabular}[t]{c}$\lambda = 20\text{cm}$\end{tabular}}}}%
    \put(0.59609571,0.23458193){\makebox(0,0)[t]{\lineheight{1.25}\smash{\begin{tabular}[t]{c}$\lambda = 13.3\text{cm}$\end{tabular}}}}%
    \put(0.82711109,0.23458195){\makebox(0,0)[t]{\lineheight{1.25}\smash{\begin{tabular}[t]{c}$\lambda = 6.6\text{cm}$\end{tabular}}}}%
    \put(0.96053275,0.20911154){\makebox(0,0)[t]{\lineheight{1.25}\smash{\begin{tabular}[t]{c}$+$\end{tabular}}}}%
    \put(0.99609938,0.11291971){\rotatebox{90}{\makebox(0,0)[t]{\lineheight{1.25}\smash{\begin{tabular}[t]{c}$\text{Re}(\hat{I})$\end{tabular}}}}}%
    \put(0.96053275,0.00592314){\makebox(0,0)[t]{\lineheight{1.25}\smash{\begin{tabular}[t]{c}$-$\end{tabular}}}}%
  \end{picture}%
\endgroup%

    \end{small}
    \caption[Frequency-space rendering of the \texttt{Cornell Box} scene]{\textbf{Frequency-space rendering of the \texttt{Cornell Box} scene.} We show the real part of the complex-valued frequency-space rendering $\hat{I}(x, y, \Omega)$ at different modulation frequencies $\Omega = 1 / \lambda$. As the frequency increases, the phase wraps around more times. All renders use our \texttt{transient\_path} integrator from \mitransient with the \texttt{phasor\_hdr\_film} plugin.}
    %\Description{figure description}
    \label{fig:mitransient:frequency_rendering}
\end{figure}

To avoid excessive memory storage and computation times, we incrementally compute the Fourier transform of the transient image for each of the Monte Carlo samples during rendering:
\begin{equation}
    \hat{I}(x, y, \Omega) \approx \frac{1}{N} \sum_{j=1}^N \frac{\mathcal{C}(\bar{\textbf{x}}_j, \,\text{tof}(\bar{\textbf{x}}_j)) \, e^{-i \Omega \,\text{tof}(\textbf{x}_j)}}{p(\bar{\textbf{x}}_j))},
\end{equation}
as an alternate version of \eref{eq:mitransient:monte-carlo-estimator}, which also yields an unbiased estimate. Finally, we showcase the resulting frequency-space render in \fref{fig:mitransient:frequency_rendering} in the \texttt{Cornell Box}. The scene is approximately 1 meter wide as reference, and we show frequencies $\Omega = 1 / \lambda$ in the range of 5 to 16.6 $\text{m}^{-1}$ which correspond to wavelengths $\lambda$ in the range of $0.06$ to $0.2 \text{m}$. As the frequencies increase, the phase wraps around more times.

%%%%%%%%%%%%%%%%%%%%%%%%%%%%%%%%%%%%%%%%%%%%%%%%%%%%%%%%%%%%%%%%%%%%%%%%%%%%%%%%%%%%%%%%%%%%%%%%%%%%%%%%
%%%%%%%%%%%%%%%%%%%%%%%%%%%%%%%%%%%%%%%%%%%%%%%%%%%%%%%%%%%%%%%%%%%%%%%%%%%%%%%%%%%%%%%%%%%%%%%%%%%%%%%%
%%%%%%%%%%%%%%%%%%%%%%%%%%%%%%%%%%%%%%%%%%%%%%%%%%%%%%%%%%%%%%%%%%%%%%%%%%%%%%%%%%%%%%%%%%%%%%%%%%%%%%%%
\subsection{Differentiable transient rendering}
\label{sec:mitransient:differentiable}

\begin{figure}
    \centering
    \captionsetup{skip=-6pt}
    \def\svgwidth{\columnwidth}
    \begin{small}
        %% Creator: Inkscape 1.2.2 (b0a8486541, 2022-12-01), www.inkscape.org
%% Accompanies image file 'forwarddiff.pdf' (pdf, eps, ps)
\begingroup%
  \makeatletter%
  \providecommand\color[2][]{%
    \errmessage{(Inkscape) Color is used for the text in Inkscape, but the package 'color.sty' is not loaded}%
    \renewcommand\color[2][]{}%
  }%
  \providecommand\transparent[1]{%
    \errmessage{(Inkscape) Transparency is used (non-zero) for the text in Inkscape, but the package 'transparent.sty' is not loaded}%
    \renewcommand\transparent[1]{}%
  }%
  \providecommand\rotatebox[2]{#2}%
  \newcommand*\fsize{\dimexpr\f@size pt\relax}%
  \newcommand*\lineheight[1]{\fontsize{\fsize}{#1\fsize}\selectfont}%
  \ifx\svgwidth\undefined%
    \setlength{\unitlength}{890.16412209bp}%
    \ifx\svgscale\undefined%
      \relax%
    \else%
      \setlength{\unitlength}{\unitlength * \real{\svgscale}}%
    \fi%
  \else%
    \setlength{\unitlength}{\svgwidth}%
  \fi%
  \global\let\svgwidth\undefined%
  \global\let\svgscale\undefined%
  \makeatother%
  \begin{picture}(1,0.33470823)%
    \lineheight{1}%
    \setlength\tabcolsep{0pt}%
    \put(0,0){\includegraphics[width=\unitlength,page=1]{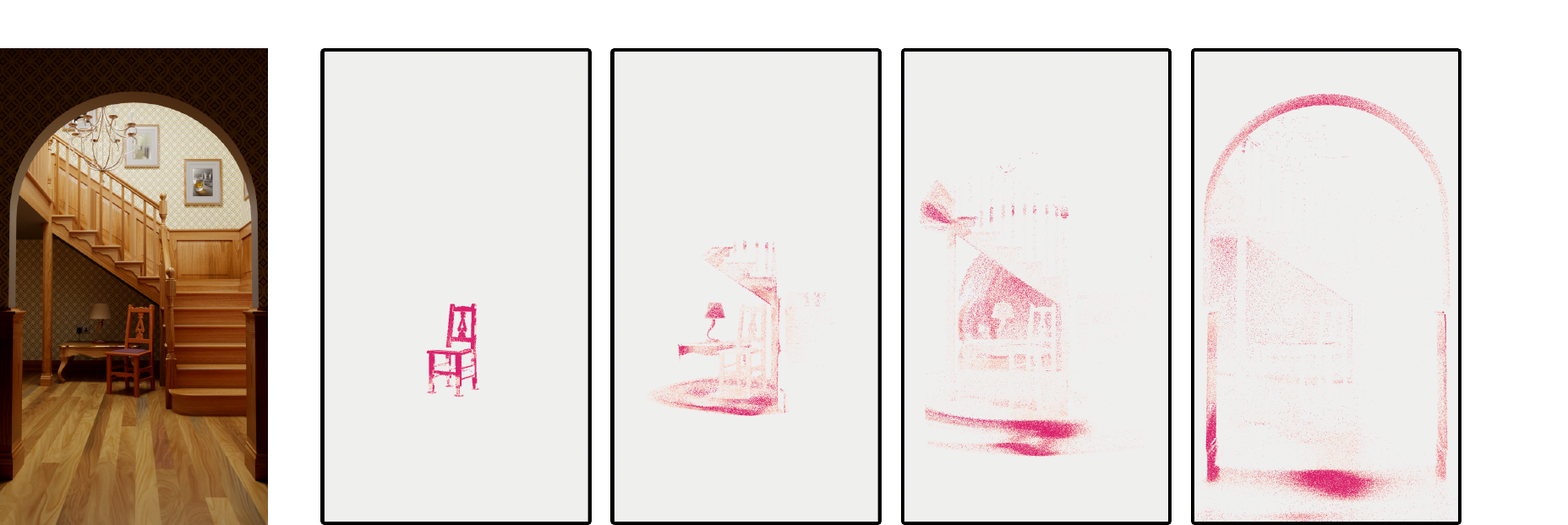}}%
    \put(0.08803592,0.31613786){\color[rgb]{0,0,0}\makebox(0,0)[t]{\lineheight{1.25}\smash{\begin{tabular}[t]{c}Steady state \end{tabular}}}}%
    \put(0.20528104,0.31613787){\color[rgb]{0,0,0}\makebox(0,0)[lt]{\lineheight{1.25}\smash{\begin{tabular}[t]{l}Forward derivatives \end{tabular}}}}%
    \put(0,0){\includegraphics[width=\unitlength,page=2]{forwarddiff.pdf}}%
    \put(0.99229207,0.15180489){\rotatebox{90}{\makebox(0,0)[t]{\lineheight{1.25}\smash{\begin{tabular}[t]{c}${\partial I}/{\partial \pi_i}$\end{tabular}}}}}%
    \put(0.05245267,0.04130021){\color[rgb]{1,1,1}\makebox(0,0)[rt]{\lineheight{1.25}\smash{\begin{tabular}[t]{r}$\pi_i$\end{tabular}}}}%
    \put(0.97458354,0.0006618){\makebox(0,0)[lt]{\lineheight{1.25}\smash{\begin{tabular}[t]{l}0\end{tabular}}}}%
    \put(0.97459942,0.28336388){\makebox(0,0)[lt]{\lineheight{1.25}\smash{\begin{tabular}[t]{l}+\end{tabular}}}}%
    \put(0,0){\includegraphics[width=\unitlength,page=3]{forwarddiff.pdf}}%
  \end{picture}%
\endgroup%

    \end{small}
    \caption[Differentiable transient rendering in the \texttt{Staircase} scene (forward)]{\textbf{Forward mode transient differentation example.} Forward mode differentiation allows to compute the derivatives of the rendered transient video $I$ with respect to one parameter $\pi_i$ of the scene. Here, we show the derivatives of $I(x, y, t)$ at different instants $t$ in \emph{world time}, with respect to the roughness of the chair $\pi_i$.}
    \label{fig:mitransient:differentiable-forward}
\end{figure}

Differentiable rendering allows tracking derivatives through the rendering computations, enabling a wide variety of inverse rendering and optimization applications. Our tool \mitransient\ supports forward and backward mode differentiation. To understand the difference, consider $\pi$ to be the set of scene parameters (e.g., material properties, albedos...). We extend \eref{eq:mitransient:transient-path-integration} to explicitly note that the rendered image $I_\pi$ and the path contribution function $\mathcal{C}_\pi$ depend on the scene parameters $\pi$ as:
\begin{equation}
    I_\pi(x, y, t) = \int_{\mathcal{X}(x, y)} \mathcal{C}_\pi(\bar{\textbf{x}}, t) \, d\mu(\bar{\textbf{x}}).
\end{equation}
Using this formulation, we can define two differentiation modes:

\emph{Forward} mode differentiation allows you to compute $\partial I / \partial \pi_i$ for one subset of parameters $\pi_i \subseteq \pi$ of the scene. This mode is particularly efficient when the number of inputs $\pi_i$ is very small, and the dimensionality of the output is large (which is the case for $\partial I / \partial \pi_i$). We showcase this mode in \fref{fig:mitransient:differentiable-forward}, where we compute the time-resolved derivatives of the rendered image $I$ with respect to one parameter $\pi_i$ that represents the roughness of the central chair. We show the derivatives in \emph{world time}, the first frame shows the chair itself, while the later frames show indirect reflections of the chair on the walls and floor as they propagate through the scene.

\begin{figure}
    \centering
    \captionsetup{skip=-4pt}
    \def\svgwidth{\columnwidth}
    \begin{small}
        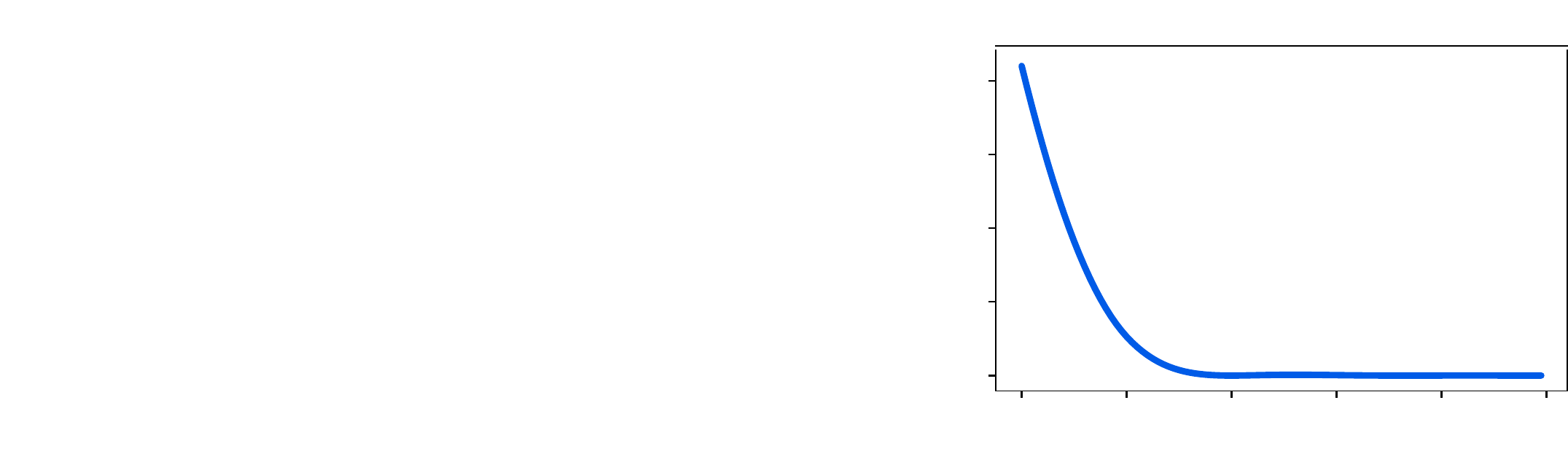
    \end{small}
    \caption[Differentiable transient rendering in the \texttt{Cornell Box} scene (backward)]{\textbf{Transient inverse rendering example.} We optimize the index of refraction $\pi_i$ of the left cube in the \texttt{Cornell Box} scene using backward mode differentiation. In the middle, we show how the transient aspect of our differentiable rendering pipeline highlights the difference between the target $I^\ast$ and our initial estimated $I_\pi$ images, as the index of refraction $\pi_i$ changes the time of flight through the cube. We optimize $\pi_i$ according to a loss function $\mathcal{L}(I_\pi, I^\ast) = \Sigma (I_\pi - I^\ast)^2$.}
    \label{fig:mitransient:differentiable-backward}
\end{figure}

\emph{Backward} mode differentiation, while it can conceptually compute the same values as forward mode, is better suited for inverse problems. For example, consider that you want to adjust a subset of the scene parameters $\pi_i \subseteq \pi$ such that the render $I_\pi$ matches a target $I^\ast$:
\begin{equation}
    \pi_i^\ast = \arg \min_{\pi_i} \mathcal{L}(I_\pi, I^\ast),
    \label{eq:loss-diff}
\end{equation}
according to a loss function $\mathcal{L}$. Contrary to forward mode, backward mode differentiation is efficient when the number of outputs is small (e.g., in \eref{eq:loss-diff} this is just a scalar loss), and the number of inputs (the parameters $\pi_i$) is large. Here, backward mode differentiation allows you to compute $\nabla_{\pi_i} \mathcal{L}$ for a subset $\pi_i \subseteq \pi$ of the scene parameters for optimizations:
\begin{equation}
    \pi_i^{s+1} \leftarrow \pi_i^s - \alpha \nabla_{\pi_i}\mathcal{L}(I_{\pi^s}, I^\ast),
\end{equation}
with a learning rate $\alpha$, updating the scene parameters $\pi_i^s$ at step $s$. We showcase this mode in \fref{fig:mitransient:differentiable-backward}, where we optimize the index of refraction of the left cube in the the \texttt{Cornell Box} scene. Note that, contrary to conventional differentiable rendering, the loss function in \eref{eq:loss-diff} has much more information through the temporal dimension of the render $I$ (see, e.g., how the differences in the middle comparison of \fref{fig:mitransient:differentiable-backward} are much more noticeable in the transient domain, as the time of flight of light varies with the index of refraction $\pi_i$). We use a simple squared error loss $\mathcal{L}(I_\pi, I^\ast)$ to optimize the index of refraction $\pi_i$ using gradient descent to match the original image $I^\ast$.

As a result, backward mode differentiation in \mitransient\ enables a plethora of physically-based forward models that can be directly plugged in into computational imaging applications as future work. For more details on how to use our transient differentiable rendering features, you can check our tutorials for forward and backward differentiable rendering in the documentation\footnote{\url{https://mitransient.readthedocs.io/en/latest/src/tutorials/diff_tutorials.html}}.

%---------------------------------------
%---------------------------------------
%---------------------------------------
\subsection{Non-line-of-sight (NLOS) scene capture simulation}
\label{sec:nlos}

Non-line-of-sight imaging refers to the family of algorithms that reconstruct objects hidden from direct view by analyzing time-resolved indirect illumination on a visible wall.

Conventional path-tracing techniques are ill-suited for NLOS scenes, as most of the signal that is relevant for this problem comes from very specific multiple-bounce paths which the algorithm has a very low probability to find.
For this purpose, we have added custom sampling techniques tailored to NLOS scenes~\cite{royo2022non}, which we provide in the \texttt{\textbf{transient\_nlos\_path}} plugin.

\begin{figure}[t]
    \centering
    \captionsetup{skip=9pt}
    \def\svgwidth{0.9\columnwidth} 
    \begin{small}
    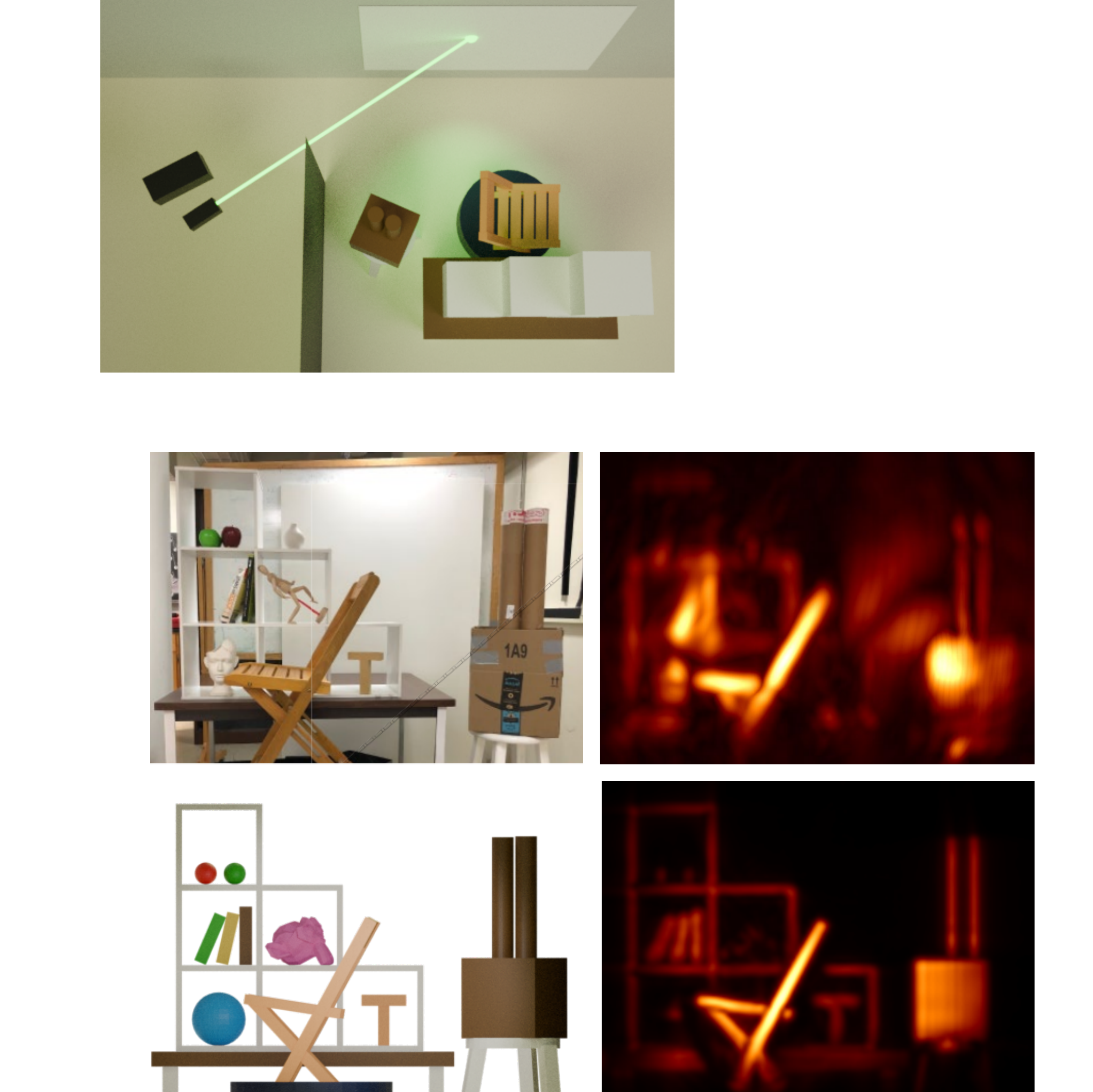
    \end{small}
    \caption{
     In order to showcase our system in a NLOS scenario, we have recreated the \textsc{Office} scene from Liu~et~al.~\cite{Liu2019phasor}. \textbf{(a)}~The laser emits a pulse towards the relay wall, and the ultra-fast camera captures the response of the hidden scene. We show one pixel of this time-resolved response. \textbf{(b)}~NLOS reconstructions of the captured and synthetic scenes.
    }
    \label{fig:officescene}
\end{figure}

To showcase the potential of our work in this area, we have recreated the setup from Liu~et~al.~\cite{Liu2019phasor}, as seen in \fref{fig:officescene}a. 
The hidden scene consists of shelves, a wooden chair, and a cardboard box. 
Similar to the original experiment, we compute the impulse response of the hidden scene on a $180 \times 130$ grid of points on the relay wall, and as can be seen in \fref{fig:officescene}b, our system can be used for NLOS reconstructions.

Our sampling techniques greatly improve the convergence time: this experiment requires only four minutes of execution time on an Intel Xeon E5-2697 CPU using 500.000 samples per pixel, which would take hours otherwise.
Nevertheless, using 5.000 samples (with four seconds of execution time) gives almost the same result.

% In \fref{fig:officescene}b, we show the reconstructions obtained using existing NLOS imaging algorithms \cite{Liu2019phasor}.
% Finally, in \fref{fig:officescene}c we show examples of the resulting signal-to-noise ratio by varying the number of random samples used by our algorithm. Note that the hardware noise models from \sref{sec:noise} can also be applied in this case.

\paragraph{\texttt{y-tal}: A software toolkit for NLOS captures.}
%
% Mitsuba 3 offers many more degrees of customization than what is usually required for quick prototyping of NLOS imaging algorithms. For this purpose, we also make available the \texttt{y-tal} Python library and command-line utility, which offers a simple interface to interact with Mitsuba 3 tailored for NLOS setups. It is publicly available on GitHub\footnote{\url{https://github.com/diegoroyo/tal/}}, and through PyPi\footnote{\url{https://pypi.org/project/y-tal/}}.
%
% Shorter version
% ---
We also make available a Python library and command-line utility, which offers a simple interface to interact with our system tailored for NLOS setups.
It is publicly available on GitHub\footnote{\url{https://github.com/diegoroyo/tal/}}, and through PyPi\footnote{\url{https://pypi.org/project/y-tal/}}.
\paragraph{Realistic hardware noise.}
\label{sec:noise}
Light transport simulations ignore many sources of noise that happen using real hardware (e.g. temporal jittering of the signal~\cite{hernandez2017computational}). 
To help bridge the gap between perfect simulations and the real world, our system can also use measures from real hardware devices to process the signal, adding realistic noise.

\begin{acks}
The authors of this work thank multiple funding sources. Diego Royo was supported by a Gobierno de Aragón predoctoral grant (CUS/803/2021) and the Gobierno de Aragón's Departamento de Ciencia, Universidad y Sociedad del Conocimiento through the Reference Research Group "Graphics and Imaging Lab" (ref T34\_23R). Jorge Garcia-Pueyo was supported by the FPU23/03132 predoctoral grant and from the European Commission’s Horizon Europe Research and Innovation Actions project Sestosenso under GA number 101070310. Óscar Pueyo-Ciutad was supported by the FPU22/02432 predoctoral grant. Óscar Pueyo-Ciutad, Guillermo Enguita and Diego Bielsa were supported by European Union’s European Defense Fund under grant agreement No 101103242.
\end{acks}

\bibliographystyle{ACM-Reference-Format}
\bibliography{bibliography}

\end{document}